\documentclass[a4paper]{jpconf}
\usepackage{graphicx}
\begin{document}
\title{(Super-)Gravities of a different sort}

\author{Jos\'e D. Edelstein$^{\dag\, 1,2,3}$ and Jorge Zanelli$^{*\, 1}$}

\address{$^1$ Centro de Estudios Cient\'{\i}ficos (CECS), Casilla 1469, Valdivia,
Chile}
\address{$^2$ Departamento de F\'\i sica de Part\'\i culas, Universidade
de Santiago de Compostela and Instituto Galego de F\'\i sica de Altas Enerx\'\i
as, E-15782 Santiago de Compostela, Spain}
\address{$^3$ Instituto de F\'\i sica de La Plata (IFLP), Universidad
Nacional de La Plata, Argentina}

\ead{$^{\dag}$jedels[at]usc[dot]es, $^{*}$jz[at]cecs[dot]cl}
\begin{abstract}
We review the often forgotten fact that gravitation theories invariant under
local de Sitter, anti-de Sitter or Poincar\'{e} transformations can be
constructed in all odd dimensions. These theories belong to the Chern--Simons
family and are particular cases of the so-called Lovelock gravities,
constructed as the dimensional continuations of the lower dimensional Euler
classes. The supersymmetric extensions of these theories exist for the AdS and
Poincar\'e groups, and the fields are components of a single connection for the
corresponding Lie algebras. In 11 dimensions these supersymmetric theories are
gauge theories for the $osp(1|32)$ and the \emph{M} algebra, respectively. The
relation between these new supergravities and the standard theories, as well as
some of their dynamical features are also discussed.
\end{abstract}


\section{Gravity as a gauge theory} 

In 1915, Einstein invented the first nonabelian gauge theory \cite{Einstein},
although at the time nobody --certainly not Einstein himself-- had any clue of
this fact, and much less, of its importance. It took some forty years and the
discovery of Yang and Mills \cite{Yang-Mills} to conceive gravitation as a
nonabelian gauge theory \cite{Utiyama,Kibble}. The original observation that
led Einstein to General Relativity (\textbf{GR}) was that the content of the
Equivalence Principle is the possibility of retaining the Lorentz symmetry of
Special Relativity in every local neighborhood of a curved spacetime. This
turns the global $SO(3,1)$ symmetry of Special Relativity into a local (gauge)
symmetry in GR.

Local Lorentz invariance is an exact gauge symmetry of GR, closely related to
the gauge symmetries that characterize the other forces of nature. In spite of
this formal similarity between gravity and the other fundamental forces of
nature, there exist a number of differences, which may be at the root of the
obstructions towards the quantum description of gravitational phenomena.


The principle of equivalence states that spacetime is a differentiable
pseudo-Riemannian manifold $M$, endowed with a tangent bundle of flat Minkowski
spaces at each point. Thus, spacetime is the base manifold for a fiber bundle
where each fiber is the Lorentz group. Note that the local Lorentz symmetry is
unrelated to the freedom to make arbitrary coordinate choices on M
--diffeomorphism invariance or general covariance. Coordinates are just
auxiliary labels and, as such, any well-posed description of the physical world
must be insensitive to the choice of coordinates. General covariance is neither
an exclusive feature of GR, nor is it a useful physical symmetry. Proving
invariance of a physical system under coordinate transformations is as
fundamental as proving invariance of ideas under a change of printer's font.

A more compelling reason to avoid using general covariance as a symmetry
principle is the fact that its first class generators do not form a Lie algebra
but an \emph{open} algebra, where the analogues of the structure constants are
functions of the phase space variables instead of being invariants under the
action of the group \cite{Henneaux85}.


Invariance of gravity under local Lorentz transformations is manifest when one
writes the Einstein--Hilbert action in the first order formalism. For example,
in four dimensions
\begin{equation}\label{EH4D}
    I_{EH}=\int d^4x\sqrt{-g}\, R = \int \epsilon_{abcd}\, R^{ab} \wedge
    e^c \wedge e^d ~.
\end{equation}
Here $R^{ab}=d\omega^{ab} + \omega^a_{\;\;c}\wedge \omega^{cb}$ is the
curvature two-form, $\omega^{ab}$ is the Lorentz (spin) connection, and $e^a =
e^a_{\;\mu}\, dx^{\mu}$ is the local orthonormal frame (vierbein). Under a
local Lorentz transformation $\Lambda^a_{\;b}(x)$, $e_a$ and $\omega^{ab}$
transform, respectively, as a vector and as a connection
\begin{eqnarray}\label{Lorentz}
    e^a &\rightarrow& e'^a(x) = \Lambda^a_{\;b}(x)\, e^b(x) ~,
    \nonumber\\[1ex]
    \omega^{ab} &\rightarrow& \omega'^{ab}(x) = \Lambda^a_{\;c}(x)\,
    \Lambda^b_{\;d}(x)\, \omega^{cd}(x) + \Lambda^a_{\;c}(x)\,
    d\Lambda^c_{\;d}(x) ~.
\end{eqnarray}
Clearly, the form of the Lagrangian (\ref{EH4D}) is quite different from the
Yang-Mills (\textbf{YM}) one, $(1/4)Tr[\textbf{F}^2]$. An obvious difference is
that (\ref{EH4D}) is linear rather than quadratic in the curvature. More
importantly, gravity requires two fields of different nature: a gauge

connection for the Lorentz group, $\omega^a_{\;\;b}$, and a vector under the
same group, $e^a$. Yang-Mills theory, on the other hand, requires no other
dynamical field but the gauge connection $\textbf{A}$. In YM theory the
spacetime metric represents a non-dynamical background of fixed geometry. On
the contrary, for a theory like Gravitation, that dynamically determines the
spacetime geometry, a prescribed background geometry would make no sense.


The two 1-form fields $e^a$, $\omega^a_{\;\;b}$, embody two essential aspects
of geometry: metricity and parallelism. These are conceptually independent
properties, the first related to the notion of distance, area, orthogonality,
and the second to the definition of parallel transport of vectors in open
neighborhoods. Since these definitions are logically independent, it is fitting
to describe them by means of dynamically independent fields. Hence, the
equivalence principle can be taken to mean that a $D$-dimensional spacetime
geometry should be described mathematically by an action principle of the form
\begin{equation}\label{GravD}
    I[e,\omega]=\int_M L_D(e,\omega,de, d\omega) ~,
\end{equation}
where the Lagrangian $L_D$ is a D-form constructed out of the fundamental
fields and their exterior derivatives. In order to ensure the Lorentz
invariance of the dynamics, it would be sufficient to require the Lagrangian to
be a Lorentz scalar. This last requirement is not strictly necessary and can be
relaxed, requesting instead that under a Lorentz transformation
(\ref{Lorentz}), the action changes by a surface term. To construct $L_D$, the
two invariant tensors of the Lorentz group,
\begin{equation}\label{inv-tensors}
    \eta_{ab}, \;\; \mbox{and} \;\; \epsilon_{a_1 a_2 \cdots a_D} ~,
\end{equation}
should also be used.


Since the metric is not a basic field in this formulation ($g_{\mu \nu}=
\eta_{ab}\, e^a_{\mu}e^b_{\nu}$), it cannot be assumed a priori to be
invertible, as in a theory defined on a prescribed background geometry. In
fact, it is conceivable that the ground state (vacuum) of gravity may
correspond to a configuration with $e^a_{\mu}=0$ \cite{Witten}. Thus, it would
be inconsistent to introduce a structure like the Hodge dual (*) which requires
the existence of a metric and its inverse, $g^{\mu \nu}$. The absence of the
*-dual does not represent a limitation since all gravity theories which yield
second order field equations for the metric can be obtained in this way.
Indeed, by taking exterior derivatives of $e^a$ and $\omega^a_{\;\;b}$ the only
Lorentz tensors that can be produced are the curvature $R^{ab}$, and torsion
$T^a=de^a+ \omega^a_{\; \;b} \wedge e^b$ two forms. Moreover, by virtue of the
Bianchi identities,
\begin{eqnarray}\label{Bianchi}
    DR^{ab} &=& dR^{ab} + \omega^a_{\;\;c}\, \wedge R^{cb} +
    \omega^b_{\;\;c}\, \wedge R^{ac} \equiv 0 \, \nonumber\\[1ex]
    DT^a &=& R^a_{\;\;b}\, \wedge e^b ~,
\end{eqnarray}
it is clear that no new tensors can be generated. These tensors, together with
the invariants (\ref{inv-tensors}) are the only ingredients at hand to build up
all gravity actions in any dimension.

\section{Three series} 

There are relatively few Lagrangians that can be written in a given spacetime
dimension $D$, with the ingredients listed above that are Lorentz invariant
$D$-forms. These candidate actions fall into three families.

\subsection{Lovelock series} 

General Relativity, viewed as a dynamical theory for the metric without
torsion, is generalized for a spacetime dimension $D>4$, by the so-called
Lovelock theories of gravity \cite{Lovelock,Zumino}. Their Lagrangians are the
most general $D$-forms built out of $R^{ab}$ and $e^a$. They take the form
\footnote{Hereafter, wedge product between forms is always understood.}
\begin{equation}\label{Lovelock1}
    L = \sum_{p=0}^{n} \alpha_p\, L_{(p)} ~,
\end{equation}
where $n = [(D-1)/2]$, $\alpha_p$ are arbitrary coefficients and
\begin{equation}\label{Lovelock2}
    L_{(p)} = \epsilon_{a_1 a_2 \cdots a_D}\, R^{a_1a_2} \cdots
    R^{a_{2p-1}a_{2p}}\, e^{a_{2p+1}} \cdots e^{a_D} ~.
\end{equation}
Included in the series (\ref{Lovelock1}) are the cosmological constant term
$L_{(0)}=\epsilon_{a_1 a_2 \cdots a_D}\, e^{a_1} \cdots e^{a_D}$, the
Einstein--Hilbert density $L_{(1)} = \epsilon_{a_1 a_2 \cdots a_D}\,
R^{a_1a_2}\, e^{a_3} \cdots e^{a_D}$, the Gauss-Bonnet density $L_{(2)} =
\epsilon_{a_1 a_2 \cdots a_D}\, R^{a_1a_2}\, R^{a_3a_4}\, e^{a_5} \cdots
e^{a_D}$, etc. For even $D$, the last term is the $D$-dimensional Euler density
$L_{(D/2)} = \epsilon_{a_1 a_2 \cdots a_D}\, R^{a_1a_2}\, R^{a_3a_4} \cdots
R^{a_{D-1}a_D}$. Each $L_{(p)}$ corresponds to the dimensional continuation to
$D$ dimensions of the $2p$-dimensional Euler density.

Varying the action with respect to the vielbein yields a generalization of
Einstein equations for arbitrary dimensions known as Lovelock equations. The
variation with respect to $\omega^{ab}$ yields an equation involving torsion
which is always solved by $T^a = 0$, but this is not the most general solution.

\subsection{Torsional series} 

Torsion is not included in the Lagrangian (\ref{Lovelock1}), although it is not
set identically equal to zero. This means that including torsion in the
Lagrangian is as legitimate as including curvature. This means that there is a
series of Lorentz invariant polynomials, which are not included in the Lovelock
series that can be added in each dimension
\begin{equation}\label{Torsional}
    L^{\rm Tor}_D(e^a,\omega^{ab}) = \sum_s \beta_s\, P^s(e^a,R^{ab},T^a) ~.
\end{equation}
Notice that these polynomials cannot involve the totally antisymmetric symbol
$\epsilon_{abc\cdots}$. The explicit form of these terms is not very
illuminating and takes a different form in each dimension. The construction of
these polynomials, as well as a broad discussion about them, were given in
\cite{Mardones-Zanelli}. These polynomials include the Pontryagin invariant
$4k$-forms, $\mathcal{P}_{4k}(R) = R^{a_1}_{\;a_2} \cdots R^{a_{2k}}_{\;a_1}$,
as particular cases.

There exist two additional terms in this family that can be included in four
dimensions, $t = T^a T_a$ and $r = R^{ab} e_a e_b$. It turns out, however, that
the combination $N_4\equiv t-r$ is a total derivative (the Nieh--Yan invariant)
and hence the two terms are equivalent Lagrangians. This type of invariants are
also related to Chern-Pontryagin classes, and may also contribute to the chiral
anomaly in spacetimes with torsion
\cite{Chandia-Zanelli1,Chandia-Zanelli2,Chandia-Zanelli3}.

\subsection{Lorentz CS series} 

There is another class of actions that are not exactly invariant, but
\emph{quasi-invariant}, under local Lorentz transformations. These are the
Lorentz Chern--Simons (\textbf{LCS}) forms that exist for all odd dimensions.
The simplest LCS form in three dimensions is $L^{CS}_3 = \omega^a_{\;b}
d\omega^b_{\;a} + \frac{2}{3} \omega^a_{\;b} \omega^b_{\;c} \omega^c_{\;a}$,
and, in general, in $4k-1$ dimensions, it takes the form
\begin{equation}\label{LCS-3}
    L^{\rm CS}_{4k-1} = [\omega (d\omega)^{2k-1}] +\gamma_1\, [\omega^3
    (d\omega)^{2k-2}] + \gamma_2\, [\omega^5 (d\omega)^{2k-3}] + \cdots +
    \gamma_{2k-1}\, [\omega^{4k-1}] ~,
\end{equation}
where the coefficients $\gamma_s$ are fixed rational numbers and the bracket
$[\cdots]$ denotes a trace. These forms yield Lorentz invariant equations
despite the fact that they involve explicitly the connection and are not truly
Lorentz invariants. This "miracle" stems from the fact that the exterior
derivative of CS forms are topological invariant densities. In this case,
$dL^{CS}_{4k-1}=\mathcal{P}_{4k}(R)=[R^{2k}]$.

\section{Miraculous choice}  

The number of Lagrangians generated in this way increases with dimension, and
so does the number of arbitrary dimensionful coupling constants
$\alpha_1,\cdots \alpha_n; \beta_1 \cdots \beta_m$. Fortunately, for $D=4$
there is complete agreement with GR: in the absence of torsion, only the
Einstein-Hilbert and the cosmological constant term can be present. What is the
meaning of all those coupling constants in higher dimensions? It can be easily
seen that they represent the existence of several scales in the theory, which
are related to different radii of curvature in the solutions, or to different
cosmological constants $\lambda_1,\lambda_2,\cdots \lambda_s$. If the
cosmological constant is a problem in four dimensions, the problem is a priori
much worse for $D>4$. It can be seen that field equations admit spacetime
geometries that jump discontinuously from one with $\lambda = \lambda_1,$ to
another with $\lambda = \lambda_2$ \cite{Boulware-Deser,Teitelboim-Zanelli}.

The presence of so many dimensionful constants endows the theory with a bad
prospect for its quantization. How could the theory be protected from
uncontrollable ultraviolet divergences? The ideal situation is closer to the
opposite extreme: a Lagrangian with no arbitrary dimensionful constants. That
is the case of a Chern--Simons theory, in which all constants are fixed
dimensionless rational numbers.


The good news is that in every dimension there exists a choice of coefficients
$\alpha_1 \cdots \alpha_n; \beta_1 \cdots \beta_m$ such that all cosmological
constants are the same and therefore there is only one scale in the theory. In
odd dimensions, this choice is even more miraculous since all dimensionful
coefficients in the action can be absorbed by means of a rescaling of the
vielbein, $e^a \rightarrow l^{-1} e^a$. Indeed, choosing the Lovelock
coefficients in (\ref{Lovelock1}) as
\begin{equation}\label{alphas}
    \alpha_p = \frac{l^{2p-D}}{D-2p} \left(
    \begin{array}{c} \frac{D-1}{2} \\ p \end{array}
    \right) ~,
\end{equation}
produces a Lagrangian that describes a theory of gravity with no built-in
scale, being, therefore, scale-invariant. This choice has an additional bonus
feature because the gauge symmetry is now enlarged from the Lorentz to the AdS
group.


As it is well known, miracles don't exist; they are either hoaxes or surprises
from our poor understanding of things. All the miracles that come with the
choice (\ref{alphas}) are consequence of the fact that with this choice the
vielbein and the Lorentz connection are combined into a connection for the AdS
group. In other words, the gauge group $SO(D-1,1)$ has been embedded into
$SO(D-1,2)$, in the form
\begin{equation}\label{AdSconnection}
    {\bf A} = \frac{e^a}{l} J_a + \frac{1}{2} \omega^{ab} J_{ab} ~,
\end{equation}
and the action becomes a functional of this connection ${\bf A}$, and not a
functional of $e^a$ and $\omega^{ab}$ separately. The Lagrangian can now be
expressed as
\begin{equation}\label{CS}
    L^{\rm CS}_{2n-1}({\bf A}) = \kappa < {\bf A} (d{\bf A})^{n-1} +
    \gamma_1\, {\bf A}^3 (d{\bf A})^{n-2} \cdots \gamma_{n-1}\,
    {\bf A}^{2n-1} > ~,
\end{equation}
where $<\cdots>$ denotes a trace on the matrix representation of the AdS
generators $J_a$, $J_{ab}$ \cite{Chamseddine1,JJG} (for details and a
comprehensive list of references see, e.g.,
Refs.\cite{ZanelliVdeL1,ZanelliVdeL2}).


It is also possible to construct a de-Sitter invariant action, (with $SO(D,1)$
as the gauge group) which is obtained by replacing $l^2 \rightarrow -l^2$ in
(\ref{alphas}). Finally, there is also the possibility of taking the vanishing
cosmological constant limit, $l \rightarrow \infty$, which yields a theory
invariant under the Poincar\'e group
\cite{Chamseddine2,Banados-Teitelboim-Zanelli}.

It is sometimes argued that the Einstein--Hilbert action with cosmological
constant in four dimensions provides a gauge theory for the (A)dS group,
because its dynamical fields ($e^a$ and $\omega^{ab}$) are components of the
(A)dS connection (\ref{AdSconnection}) \cite{Macdowell-Mansouri,Stelle-West}.
The problem with this point of view is that the (A)dS symmetry cannot be
respected by the action because there is no Lagrangian for the connection ${\bf
A}$ invariant under the (A)dS gauge group in four dimensions. Arguing that the
symmetry is spontaneously broken is also hard to sustain since there seems to
be no regime of the theory in which the symmetry can be restored.

\section{Surface terms and transgressions}

For mathematicians, Chern--Simons forms are not natural objects. They are not
truly invariant, changing by an exact form under a gauge transformation. In
physics this is not a serious problem because exact forms in the action are
surface terms that, generically, don't affect the field equations or the
conservation laws. However, invariances ``up to surface terms" have other
important physical consequences. The value of the conserved charges, and of the
action itself can be renormalized by surface terms. This in turn affects the
definition of thermodynamic quantities like the energy and entropy of a black
hole.

On the other hand, boundary conditions sufficient to ensure that the action
attains an extremum on the classical orbits, require to supplement the action
by a surface term of a particular form. In asymptotically locally AdS spaces
(\textbf{ALADS}) the boundary term takes the form $B_{2n}[{\bf
A},\overline{{\bf A}}]$, where the field $\overline{{\bf A}}$ is only defined
at the surface of spacetime, whose r\^ole is to match the boundary conditions
under which the action is to be varied \cite{MOTZ1}. This addition cures
several problems at once: it provides a well-defined variation while, at the
same time, it renders the charges and the on-shell value of the action finite,
producing well defined thermodynamic quantities, which can also be computed by
other means \cite{BTZentropy}.  In particular, the energy of the vacuum turns
out to be exactly the Casimir energy for AdS without additional regularizations
(counterterms) or ad-hoc background subtractions \cite{MOTZ2}.

As in other cases in physics, the solution to this problem comes from the
requirement of gauge invariance. The field $\overline{{\bf A}}$ and the
boundary term $B_{2n}[{\bf A},\overline{{\bf A}}]$ correspond precisely to what
is required to turn the Chern--Simons form into a \emph{transgression form}
\cite{Nakahara,EGH,IRS}. Unlike CS forms, transgression forms depend on two
connections, ${\cal T}_{2n-1}({\bf A},\overline{{\bf A}})$. Transgressions are
gauge invariants provided ${\bf A}$ and $\overline{{\bf A}}$ transform as
connections for the same gauge group. The defining property of a transgression
is that its exterior derivative is the difference of two invariant classes, for
${\bf A}$ and $\overline{{\bf A}}$, respectively,
\[ d{\cal T}_{2n-1}({\bf A},\overline{{\bf A}}) = <F^n({\bf A})> -
<\bar{F}^n(\overline{{\bf A}})> ~. \] Thus, the transgression form can also be
written as
\[ {\cal T}_{2n-1} = L^{CS}_{2n-1}({\bf A}) -
L^{CS}_{2n-1}(\overline{{\bf A}}) + dB_{2n}({\bf A},\overline{{\bf A}}) ~, \]
where $B_{2n}({\bf A},\overline{{\bf A}})$ is defined on a local chart over the
boundary of the spacetime manifold $M$.


The presence of the second connection field $\overline{{\bf A}}$ in the
transgression form is puzzling if considered as a second dynamical field on the
same footing as ${\bf A}$. However, $\overline{{\bf A}}$ need only be defined
at the boundary of the spacetime manifold $M$; it is sufficient to define
$\overline{{\bf A}}$ on a different manifold $\overline{M}$ that shares a
common boundary with (\emph{cobordant to}) $M$, $\partial M =
\partial \overline{M}$. Then, the action principle based on the transgression form
can be written as
\begin{equation}\label{TransAction}
    I_{Trans}[{\bf A},\overline{{\bf A}}] = \int_M L_{CS}({\bf A}, d{\bf A})
    - \int_{\overline{M}} L_{CS}(\overline{{\bf A}}, d\overline{{\bf A}})
    + \int_{\partial M} B_{2n}({\bf A},\overline{{\bf A}}) ~.
\end{equation}
The main advantage of this expression is that it allows to compute the
conserved charges by direct application of Noether's theorem in covariant
language and without subtractions \cite{Mora,MOTZ3}.

\section{Supersymmetric extensions}  

According to the Haag--Lopuszanski--Sohnius theorem
\cite{Haag-Lopuszanski-Sohnius}, supersymmetry is essentially the only
nontrivial way to extend a spacetime symmetry, circumventing the well known
obstruction pointed out by Coleman and Mandula \cite{Coleman-Mandula}. The
question then naturally arises, whether there exist supersymmetric extensions
for the theories described here. The answer is in the affirmative in the
odd-dimensional CS-AdS theories. Moreover, these theories admit a supersymmetry
which is realized like in any standard non-abelian gauge theory, namely, the
dynamical field is a connection for the (super)group and the action turns out
to be invariant off-shell (up to surface terms).

In three dimensions, the standard Einstein-Hilbert plus cosmological constant
action is a Chern--Simons theory and its supersymmetric extension has been
known for almost twenty years \cite{Achucarro-Townsend}. The resulting 2+1 AdS
supergravity is a gauge theory for the group $OSp(p|2;R) \otimes OSp(q|2;R)$.
In five dimensions, the locally supersymmetric extension of gravity was found
by Chamseddine \cite{Chamseddine2}, and its purely gravitational sector is the
CS-AdS action described above. The generalization to higher dimensions was
found in \cite{Troncoso-Zanelli1a,Troncoso-Zanelli1b}, and the supersymmetric
extensions of the Poincar\'e theory was presented in
\cite{Banados-Troncoso-Zanelli}.


The idea is to extend the action by introducing all the necessary fields to
produce a connection for the gauge supergroup that contains $AdS_D$ as a
subgroup in a given dimension. This can be done from first principles, if one
has an {\it a priori} knowledge of what are the semisimple superalgebras
containing $AdS_D$. Alternatively, one can start by adding to
(\ref{AdSconnection}) the supersymmetry generators $Q$ and $\bar{Q}$, with the
corresponding gauge fields $\bar{\psi}$ and $\psi$,
\begin{equation}\label{superAdSconnection}
    {\bf A} = \frac{e^a}{l} J_a + \frac{1}{2} \omega^{ab} J_{ab} +
    \bar{Q} \psi + \bar{\psi} Q + \cdots ~,
\end{equation}
and subsequently check the closure of the extended algebra. This requires, in
general, extra bosonic generators and, in some cases, several copies of the
fermions (this is what the dots mean in (\ref{superAdSconnection})). The result
is quite unique. It is summarized in the next table, where the field content of
the resulting theories for $D=5,7,11$, and the corresponding algebras, are
confronted with the standard supergravities.

\begin{center}
$$\begin{array}{|c|c|c||c|}
\hline
D & {\rm CS-AdS~supergravity} & {\rm Algebra} & {\rm Standard~supergravity}\\
\hline\hline
& & & \\[-1ex]
5 & e_{\mu}^a\; \omega_{\mu }^{ab}\; A_{\mu}\; \psi_{\mu}^{\alpha}\;
\bar{\psi}_{\alpha \mu} & usp(2,2|1) & e_{\mu}^{a}\;\psi_{\mu}^{\alpha}\;
A_{\mu}\;\bar{\psi}_{\alpha \mu} \\[1ex]
\hline
& & & \\[-1ex]
7 & e_{\mu}^a\; \omega_{\mu}^{ab}\; A_{j\mu}^i\; \psi_{\mu}^{i\alpha},\;
i,j=1,2 & osp(2|$8$)  & e_{\mu}^a\;A_{[3]}\; A_{j\mu}^i\; \lambda^{\alpha}\;
\phi \; \psi_{\mu}^{i\alpha},\; i,j=1,2 \\[1ex]
\hline
& & & \\[-1ex]
11 & e_{\mu}^a \;\omega_{\mu}^{ab}\; b_{\mu}^{abcde}\; \psi_{\mu}^{\alpha}\;
& osp(32|1)& e_{\mu}^a\; A_{[3]}\; \psi_{\mu}^{\alpha} \\[1ex]
\hline
\end{array}$$
\end{center}


\noindent Some general comments are in order at this point (for a detailed
discussion, see \cite{ZanelliVdeL1,ZanelliVdeL2}):

~

\noindent \textbf{Supergravities.} The actions obtained in this way are, by
construction, invariant under the gauge superalgebra and diffeomorphisms. Since
they include gravity, they are supergravities, albeit of a different sort. Some
authors would reserve the word \emph{supergravity} for supersymmetric theories
whose gravitational sector is described by the Einstein--Hilbert Lagrangian.
This narrow definition is correct in three and four dimensions, but  seems
unwarranted for $D>4$ in view of the numerous possibilities beyond EH. If one
wishes to be precise, the supergravities described here seem to belong to a
separate class and the connection with the standard ones is still an open
problem.

~

\noindent \textbf{Local supersymmetry.} The supersymmetry transformations are
those of a connection, namely, $\delta {\bf A} = -\nabla {\bf \Lambda} =
-(d{\bf \Lambda} + [{\bf A},{\bf \Lambda}])$, where ${\bf \Lambda}$ is a
zero-form with values in the Lie algebra, and $\nabla$ is the exterior
covariant derivative in the representation of ${\bf A}$. In particular, under a
supersymmetry transformation, ${\bf \Lambda} = \bar{\epsilon}^{i} Q_{i} -
\bar{Q}^{i} \epsilon_{i}$. For instance, in terms of the component fields of
the five dimensional $usp(2,2|1)$ theory, this means
\[ \begin{array}{lll}
\delta e^{a} & = & \frac{1}{2}\left( \overline{\epsilon }^{r}{\bf \Gamma}
^{a}\psi _{r}-\bar{\psi}^{r}{\bf \Gamma }^{a}\epsilon _{r}\right) ~, \\[1ex]
\delta \omega ^{ab} & = & -\frac{1}{4}\left( \bar{\epsilon}^{r}{\bf \Gamma }
^{ab}\psi _{r}-\bar{\psi}^{r}{\bf \Gamma }^{ab}\epsilon _{r}\right) ~, \\[1ex]
\delta A_{\,s}^{r} & = & -i\left( \bar{\epsilon}^{r}\psi_{s}-\bar{\psi}
^{r}\epsilon _{s}\right) ~, \\[1ex]
\delta \psi _{r} & = & -\nabla \epsilon _{r} ~, ~~~~~~~
\delta \bar{\psi}^{r} = -\nabla \bar{\epsilon}^{r} ~, \\[1ex]
\delta A & = & -i\left( \bar{\epsilon}^{r}\psi _{r}-\bar{\psi}^{r}\epsilon
_{r}\right) ~.
\end{array} \]
where $\nabla$ is the covariant derivative on the bosonic connection,
\[ \nabla \epsilon_{r} = \left( d +\frac{1}{2}\omega^{ab}{\bf \Gamma}_{ab} +
\frac{1}{2l} e^{a} {\bf \Gamma}_{a} \right) \epsilon_{r} - A_{\,r}^{s}
\epsilon_{s} - \frac{3i}{4}\, A\, \epsilon_{r} ~. \]

~

\noindent \textbf{Off-shell symmetry.} These actions are invariant (up to
surface terms) under these transformations, and neither on-shell conditions nor
auxiliary fields are necessary to realize the symmetry. This is in contrast
with the standard cases, which often require torsional on-shell conditions in
order to close the symmetry algebra. Symmetries requiring on-shell conditions
are likely to be troublesome, since they are not necessarily respected in the
quantum theory.

~

\noindent \textbf{Extended susy.} These algebras allow for extensions with
$\mathcal{N}>1$, and the field multiplets for these algebras can be easily
constructed in all cases. These algebras possess a periodic nature which is
inherited from the well known periodicity \emph{mod} $8$ of the Clifford
algebras. Thus, the relevant groups are $OSp(\mathcal{N}|2^{4k+1})$ for
$D=8k+3$, $OSp(2^{4k-1}|\mathcal{N})$ for $D=8k-1$, and
$SU(2^k,2^k|\mathcal{N})$ for $D=4k+1$.

~

\noindent \textbf{Odd $D$, $\Lambda<0$.} No similar construction is known for
positive cosmological constant. The reason is that the de Sitter group does not
admit supersymmetric extensions \cite{Sohnius,vH-VP}. Chern--Simons actions
exist only in odd dimensions; therefore, a similar construction does not exist
in even dimensions. It is possible, however that in even dimensions a
construction similar to the standard supergravity in four dimensions could be
carried out, where some on-shell conditions are assumed in order to close the
algebra including diffeomorphisms.

~

\noindent \textbf{Matching of degrees of freedom.} In these theories there is
no matching between bosonic and fermionic degrees of freedom. The matching
present in standard supersymmetric theories results from two assumptions which
do not hold in the present case: the first assumption is that the spacetime
symmetry group is Poincar\'e, while in our case it is the AdS group. The second
is that the fields form a vector multiplet under supersymmetry, which is not
the case here since all the fields are parts of a connection and therefore
belong to the adjoint representation. It is worth mentioning that global issues
--like the presence of a deficit angle-- may also spoil the boson--fermion
degeneracy in standard supergravity \cite{witetal1,witetal2,witetal3,witetal4}.

~

\noindent \textbf{Polarization states.} All component fields in these theories
carry only one spacetime index (they are 1-forms), and they are antisymmetric
tensors of arbitrary rank under the Lorentz group ({\it i.e.},
$b_{\mu}^{ab\cdots}$). Thus, they belong to representations of the rotation
group whose Young tableaux have an arbitrarily long single column and one row
with two squares. This means that the fundamental fields in these theories
describe states of spin 2 or less, which goes in the opposite direction of the
recent interest on higher spin fields \cite{Higher-spin,Sagnotti}.

~

\noindent \textbf{Degeneracy.} Chern--Simons systems in dimensions $D \geq 5$
possess remarkable dynamical features, unexpected in a field theory but often
found in fluid dynamics. One of these features stems from the fact that the
symplectic form is a function of the connection, and its rank depends on the
configuration \cite{Banados-Garay-Henneaux1,Banados-Garay-Henneaux2}. There are
regions in phase space where the symplectic form has maximal rank
(\emph{generic} configurations), where the counting of degrees of freedom is
the usual one \cite{HTZ}. Other regions, where the rank is smaller
(\emph{degenerate} configurations), possess fewer propagating degrees of
freedom. There are even \emph{maximally degenerate} configurations, around
which the theory is topological and has no local degrees of freedom. An example
of such maximally degenerate configuration is the standard vacuum of Yang-Mills
theory, ${\bf A}=0$.

Another unexpected feature is that degenerate systems may loose degrees of
freedom in their time evolution. A simple mechanical model shows that a
degenerate system may start from a nondegenerate configuration reaching a state
where the degeneracy occurs in a finite time. There, some degrees of freedom
cease to be dynamical and become gauge coordinates. After that, those degrees
of freedom, as well as their initial data, are irreversibly lost
\cite{Saavedra-Troncoso-Zanelli}.

~

\noindent \textbf{Irregularity.} An independent issue, also present in CS
theories is the fact that the functional independence of the gauge generators
(first class constraints) may break down for certain configurations \cite{MiZ},
and a careful analysis is required in order to have a well defined canonical
formalism \cite{Miskovic-Troncoso-Zanelli1,Miskovic-Troncoso-Zanelli2}.

\section{Manifest $M$-covariant theory} 

An obvious advantage of the CS construction is the economy of assumptions. The
only information required to define the Lagrangian is the gauge (super)group
and the dimensionality of the manifold. The field content, the coupling
constants, the dynamics of the spacetime manifold, the vacuum structure, are
all outputs of the theory.

As an example, consider a CS theory for the supersymmetric extension of the
Poincar\'e group in eleven dimensions. Following the steps outlined here, one
arrives almost uniquely at a gauge invariant action for the \emph{M}-algebra
\cite{Hassaine-Troncoso-Zanelli1,Hassaine-Troncoso-Zanelli2}. The connection,
\begin{equation}\label{Mconnection}
    {\bf A}=\frac{e^a}{l}P_a +\frac{1}{2}\omega^{ab}J_{ab}+ \bar{Q}\psi +
    b^{ab} Z_{ab} + b^{abcde} Z_{abcde} ~,
\end{equation}
includes, apart from the vielbein, the Lorentz connection, and the gravitino, a
second-rank and a fifth-rank antisymmetric Lorentz tensor one-forms, $b^{[2]}$
and $b^{[5]}$. The superalgebra includes the Poincar\'e generators
($P_a,J_{ab}$), one (Majorana) supersymmetry generator $Q$ and the ``central
extensions" of the $M$-algebra, $Z_{[2]}$ and $Z_{[5]}$,
\begin{equation}\label{MAlgebra}
    \{Q_{\alpha },Q_{\beta }\}=\left( C\Gamma ^{a}\right) _{\alpha \beta}
    P_{a}+(C\Gamma ^{ab})_{\alpha \beta }Z_{ab}+(C\Gamma ^{abcde})_{\alpha
    \beta}Z_{abcde} ~.
\end{equation}
The generators $Z_{[2]},\; Z_{[5]}$ commute with all but the Lorentz
generators. The supersymmetric action is found to be
\cite{Hassaine-Troncoso-Zanelli1,Hassaine-Troncoso-Zanelli2}
\begin{eqnarray}
    L_{\alpha}^{\rm M} & = & \epsilon_{a_1 \cdots a_{11}}\, R^{a_1a_2}
    \cdots R^{a_9a_{10}}\, e^{a_{11}} - \frac{1}{3}\, R_{abc}\, \bar{\psi}\,
    \Gamma^{abc} D\psi - \frac{1}{12}\, R_{abc}\, R_{de}\, b^{abcde}
    \nonumber \\
    & & + 8\, [R^{2}\, R_{ab} - 6 (R^3)_{ab}]\,
    R_{cd} \left( \bar{\psi}\, \Gamma^{abcd} D\psi -6R^{[ab}\, b^{cd]}
    \right) ~,
\label{Action}
\end{eqnarray}
where $R_{abc} = \epsilon_{abca_1 \cdots a_8} R^{a_1a_2} \cdots R^{a_7a_8}$,
$R^{2} := R^{ab}\, R_{ba}$ and $(R^{3})^{ab} := R^{ac}\, R_{cd}\, R^{db}$.

\subsection{Tentative vacuum states} 

We now turn to the dynamical contents of this system. In particular, one would
like to identify a true vacuum of the theory. The field equations take the form
\begin{equation}
    \left\langle F^{5}\, G_{A}\right\rangle =0 ~, \label{FieldEqs}
\end{equation}
where the curvature $F=dA+A^{2} = \frac{1}{2} R^{ab}\, J_{ab} + \tilde{T}^{a}\,
P_{a} + D\psi^{\alpha}\, Q_{\alpha} + \tilde{F}^{^{[2]}}\, Z_{^{[2]}} +
\tilde{F}^{^{[5]}}\, Z_{^{[5]}}$, with $\tilde{T}^{a} = De^{a} - (1/2)
\bar{\psi} \Gamma^{a} \psi$ and $\tilde{F}^{[k]} = Db^{[k]} - (1/2) \bar{\psi}
\Gamma^{[k]} \psi$ for $\,\,k=2$ and $5$. The bracket $\left\langle
...\right\rangle $ is a multilinear form of the M-algebra generators $G_{A}$.

Obviously, a configuration with a locally flat connection, $F=0$, solves the
field equations (\ref{FieldEqs}) and would be a natural candidate for the
vacuum in a standard field theory. Moreover, such state is invariant under all
gauge transformations being, therefore, maximally supersymmetric, which makes
it likely to be a stable (BPS) configuration. Identifying this solution with a
vacuum state would seem even more compelling in view of the fact that it has no
charge of any kind and is therefore invariant under all spacetime and
supersymmetry transformations.

Matter-free eleven-dimensional Minkowski spacetime is an example of such a
state. However, no local degrees of freedom propagate on such background: all
perturbations around it are zero modes. In fact, for the configuration,
$\psi=0$, $b^{[2]}=0$, $b^{[5]}=0$, Eq. (\ref{FieldEqs}) is a set of polynomial
equations of fifth degree in the curvature two-forms. In particular, the
equations obtained varying with respect to the vielbein and the spin connection
take the form
\begin{equation}\label{R5}
    \epsilon_{aa_1 \cdots a_{10}} R^{a_1a_2} \cdots R^{a_9a_{10}} = 0 ~,
\end{equation}
\begin{equation}\label{R4T}
    \epsilon_{aba_1 \cdots a_9} R^{a_1a_2} \cdots R^{a_7a_8}\, T^{a_9} = 0 ~.
\end{equation}
Thus, in order to have a propagating connection, the spatial components
$R^{ab}_{\;\;ij}$ cannot be small and must therefore be non-perturbative. Since
the derivatives of the field cannot be small either, the deviations are
necessarily non-local. In order to have well-defined linearized perturbations,
a background solution must be a simple zero of one of the set of equations. In
particular, this requires the curvature to be nonvanishing on a submanifold of
sufficiently high dimensionality.

\subsection{Nontrivial vacuum geometry} 

Let us consider a torsionless spacetime with a product geometry of the form
$X_{d+1}\times S^{10-d}$, where $X_{d+1}$ is a domain wall whose worldsheet is
a $d$-dimensional spacetime $M_{d}$. The line element is
\begin{equation}
    ds^{2} = e^{-2\xi|z|} \left(dz^{2} + \tilde{g}_{\mu \nu }^{(d)}(x)\,
    dx^{\mu }\, dx^{\nu}\right) + d\Omega_{10-d}^2 ~, \label{Ansatz}
\end{equation}
where $\tilde{g}_{\mu \nu }^{(d)}$ stands for the worldsheet metric,
$d\Omega_{10-d}^2$ is the metric of $S^{10-d}$ with radius $r_{0}$, and $\xi$
is a constant. This ansatz solves (\ref{R4T}) identically, and, as it is shown
in \cite{Hassaine-Troncoso-Zanelli1,Hassaine-Troncoso-Zanelli2}, it also solves
(\ref{R5}) and possesses propagating degrees of freedom only if $d=4$ and
$\tilde{g}_{\mu \nu }^{(d)}(x)$ describes a de Sitter geometry of cosmological
constant $\Lambda_{4}=3\xi^{2}$.

The geometry defined by (\ref{Ansatz}) is an example of a configuration with
less than the maximal number of degrees of freedom. In this case, the geometry
has the degrees of freedom of $(3+1)$-dimensional gravity with positive
cosmological constant, which is certainly less than those of the full $(10+1)$
theory. This is a generic situation among the ansatze of the form
(\ref{Ansatz}): for large enough starting dimension $D>4$, the $d$-dimensional
spacetime would have propagating degrees of freedom only if $d=4$.

\section{Gauge action for EH theory} 

As mentioned above, it is still unclear how are CS supergravities related to
the standard theories. As a step towards understanding this point, one might
look for the minimal deformation or extension of the (super) Poincar\'e group
Chern--Simons theory where the pure gravity sector is described by the
Einstein--Hilbert term. It is possible to address this problem by means of an
expansion method that allows to deform consistently a Chern--Simons theory into
another one but for a Lie (super) algebra of larger dimension
\cite{AIPV1,AIPV2,JMI}. In correspondence to the appearance of extra
generators, this requires the introduction of additional fields.

Consider the simplest case of a bosonic deformation of the Poincar\'e symmetry
in five dimensions \cite{EHTZ}. In order to cancel the variation of the
Einstein--Hilbert action, two additional bosonic fields are included, a vector
one-form $h^a$ and an antisymmetric tensorial one-form field $\kappa^{ab}$.
Consequently, the Poincar\'e connection is extended by means of two additional
generators, $Z_a$ and $Z_{ab}$,
\begin{equation}
    A = e^a P_a + \frac{1}{2} \omega^{ab}\, J_{ab} + h^a\, Z_a +
    \frac{1}{2} \kappa^{ab}\, Z_{ab} ~. \label{connection5}
\end{equation}
The resulting algebra turns out to be an extension of the Poincar\'e algebra by
an Abelian ideal spanned by these generators. The Lagrangian describing the
dynamics of the system reads
\begin{eqnarray}
    L^{\rm EH}_{\rm CS} = \epsilon_{abcdf} \Big( \frac{2}{3} R^{ab}\, e^c\,
    e^d\, e^f + R^{ab}\, R^{cd}\, h^f + 2 R^{ab}\, \kappa^{cd}\, T^f \Big) ~.
\label{action5}
\end{eqnarray}

In a torsionless, matter-free configuration ({\it i.e.}, $h^a = \kappa^{ab} =
0$), the field equations become the Einstein equations with an additional
Gauss--Bonnet constraint. This system has as a non-trivial solution a pp-wave.
Interestingly enough, allowing for $\kappa^{ab}\neq0$, the field equations
become
\begin{eqnarray}
    & & \epsilon_{abcdf} R^{ab}  e^c e^d = - \epsilon_{abcdf} R^{ab}
    D\kappa^{cd} ~, \label{eq1l} \\
    & & \epsilon_{[a\vert cdfg} R^{cd} \kappa^{fg}e_{\vert b]} = 0 ~,
    \label{eq2l} \\
    & & \epsilon_{abcdf} R^{ab}R^{cd} = 0 \label{eq3l} ~.
\end{eqnarray}
It is not hard to check that a four dimensional de Sitter domain wall,
analogous to (\ref{Ansatz}), exists if the extra bosonic field takes the form
\begin{equation}
    \kappa^{\mu\nu} = 0 ~, \qquad\qquad\qquad
    \kappa^{\mu z}=-\frac{1}{2\xi}\mbox{sgn}(z)\,e^{-2\xi\vert z\vert}
    \,\tilde{e}^{\mu} ~. \label{kappa}
\end{equation}
This is nothing but the five-dimensional version of the metric solution of the
Poincar\'e invariant gravity theory displayed above.

\section{Discussion}     

We have argued that GR represents a way to implement the Lorentz invariance at
a local level. This calls for a first order formalism, where the basic fields
are two 1-forms, $e^a$ and $\omega^a_{\;\;b}$. If we further demand an
enlargement of the gauge symmetry from the Lorentz to the Poincar\'e, dS or AdS
groups, we are pushed towards quite a unique answer: Chern--Simons (super)
gravity. These theories exist for any odd dimensional space-time. Interstingly
enough, it is possible to write down an action in eleven dimensions with the
symmetries dictated by the M algebra. This algebra, which corresponds to the
maximal extension of the ${\cal N}=1$ super Poincar\'e algebra, plays an
important r\^ole in M-theory \cite{Tow}. This is very suggestive and the
question is unavoidable: is Chern--Simons supergravity for the M--algebra
related to M--theory? Does this theory play a r\^ole in the M--theory diagram?
There have been attempts to relate these theories. It was already suggested
that M theory could be non-perturbatively equivalent to a Chern--Simons theory,
though with a different symmetry group; namely, $OSp(32|1) \times OSp(32|1)$
\cite{Horava} (see also \cite{MaxB,Nastase}). This claim was mainly supported
on arguments dealing with holography. However, the connection to eleven
dimensional supergravity at low energies, to the best of our knowledge, has not
been understood yet. We have seen that it is possible to extend the Poincar\'e
algebra in such a way that the Einstein--Hilbert action comes out. However,
several bosonic fields need to be introduced and their equations of motion
severely constrain the system. On the other hand, standard (super) gravity is
not a gauge theory of the (super) Poincar\'e group. Thus, it seems clear that
the connection between these theories possibly demand the existence of a
spontaneous symmetry breaking mechanism. One thing seems clear: a lot of
interesting results are still to be uncovered.

\ack
One of us (JZ) wishes to thank M. Hassaine, M. Henneaux, C. Mart\'{i}nez, P.
Mora, R. Olea, S. Theisen and R. Troncoso, who have taught him a great deal
through many enlightening discussions over several years. JDE is pleased to
acknowledge interesting conversations on the subject of this article with M.
Hassaine, J.M. Izquierdo and R. Troncoso. JZ wishes to thank Jeanette Nelson,
Marco Cavaglia and Mariano Cadoni for their warm hospitality and for the
delightful scientific atmosphere in Cala Gonone. This work is partially
supported by grants 3020032, 1051056 and 1020629 from FONDECYT. The work of JDE
has been supported in part by ANPCyT under grant PICT 2002 03-11624, by the FCT
grant POCTI/FNU/38004/2001, by MCyT, FEDER and Xunta de Galicia under grant
FPA2005-00188, and by the EC Commission under grants HPRN-CT-2002-00325 and
MRTN-CT-2004-005104. Institutional support to the Centro de Estudios
Cient\'{i}ficos (CECS) from Empresas CMPC is gratefully acknowledged. CECS is a
Millennium Science Institute and is funded in part by grants from Fundaci\'{o}n
Andes and the Tinker Foundation.

\section*{References}

\end{document}